\documentclass[usenatbib,useAMS]{mn2e}

\usepackage{graphicx,times}

\newcommand{\Hb}{\ensuremath{{\rm H}\beta}}
\newcommand{\Hg}{\ensuremath{{\rm H}\gamma}}
\newcommand{\HgA}{\ensuremath{{\rm H}\gamma_{\rm A}}}
\newcommand{\HgF}{\ensuremath{{\rm H}\gamma_{\rm F}}}
\newcommand{\Hd}{\ensuremath{{\rm H}\delta}}
\newcommand{\HdA}{\ensuremath{{\rm H}\delta_{\rm A}}}

\newcommand{\Mgb}{\ensuremath{{\rm Mg}\, b}}

\newcommand{\Fe}{\ensuremath{\langle {\rm Fe}\rangle}}

\newcommand{\aFe}{\ensuremath{\alpha/{\rm Fe}}}

\newcommand{\MgFep}{\ensuremath{[{\rm MgFe}]^{\prime}}}
\newcommand{\ZH}{\ensuremath{Z/{\rm H}}}

\newcommand{\JH}{\ensuremath{J\!-\!H}}

\title[Rejuvenation of spiral bulges] {Rejuvenation of spiral bulges}

\author[Daniel Thomas \& Roger L.~Davies] {Daniel Thomas$^{1,2}$ \&
Roger L.~Davies$^1$\\ $^1$University of Oxford, Denys Wilkinson
Building, Keble Road, Oxford, OX1 3RH, UK\\ $^2$Max-Planck-Institut
f\"ur extraterrestrische Physik, Giessenbachstra\ss e, D-85748
Garching, Germany}

\date{Accepted 2005 November 15. Received 2005 November 2; in original
form 2005 September 23}

\pagerange{\pageref{firstpage}--\pageref{lastpage}}

\pubyear{2006}

\begin{document}

\bibliographystyle{mn2e}
\maketitle

\label{firstpage}

\begin{abstract}
We seek to understand whether the stellar populations of galactic
bulges show evidence of secular evolution triggered by the presence of
the disc. For this purpose we re-analyse the sample of Proctor and
Sansom, deriving stellar population ages and element abundances from
absorption line indices as functions of central velocity dispersion
and Hubble type. We obtain consistent constraints on ages from the
three Balmer line indices \Hb, \Hg, and \Hd, based on stellar
population models that take the abundance ratio effects on these
indices into account. Emission line contamination turns out to be a
critical aspect, which favours the use of the higher-order Balmer line
indices. Our derived ages are consistent with those of Proctor and
Sansom based on a completely different method.  In agreement with
other studies in the literature, we find that bulges have relatively
low luminosity weighted ages, the lowest age derived being
$1.3\;$Gyr. Hence bulges are not generally old, but actually
rejuvenated systems. We discuss evidence that this might be true also
for the bulge of the Milky Way. The data reveal clear correlations of
all three parameters luminosity weighted age, total metallicity, and
\aFe\ ratio with central velocity dispersion. The smallest bulges are
the youngest with the lowest \aFe\ ratios owing to late Fe enrichment
from Type Ia supernovae. Using models combining recent minor star
formation with a base old population, we show that the smallest bulges
must have experienced significant star formation events involving
$10-30\;$ per cent of their total mass in the past $1-2\;$Gyr. No
significant correlations of the stellar population parameters with
Hubble Type are found.  We show that the above relationships with
$\sigma$ coincide perfectly with those of early-type galaxies.  In
other words, bulges are typically younger, metal-poorer and less \aFe\
enhanced than early-type galaxies, because of their smaller masses. At
a given velocity dispersion, bulges and elliptical galaxies are
indistinguishable as far as their stellar populations are concerned.
These results favour an inside-out formation scenario and indicate
that the discs in spiral galaxies of Hubble types Sbc and earlier
cannot have a significant influence on the evolution of the stellar
populations in the bulge component. The phenomenon of pseudobulge
formation must be restricted to spirals of types later than Sbc.
\end{abstract}



\section{Introduction}
\label{sec:intro}
Pseudobulges are bulges formed out of disc material in secular
processes \citep{Kormendy82}. As laid out in detail in the recent
review by \citet{KK04}, they are 'are not just dust features or the
outer disk extending inside a classical bulge all the way to the
centre', but appear to be built by nuclear star formation. In other
words, the presence of pseudobulges in spiral galaxies should be
detectable through fingerprints of relatively recent star formation in
their stellar populations. This seems to stand in clear contrast to
the commonly accepted perception that bulges are generally old
\citep{Renzini99}. There is strong and compelling evidence that the
bulk of stellar populations in the Milky Way Bulge are old without
significant amounts of recent star formation
\citep{Ortetal95,Renzini99,FWS03,Zocetal03}. On the other hand, blue
colours, patchy dust features, and low surface brightnesses are found
predominantly in bulges of later type spirals
\citep{deJong96b,Peletal99a}. Moreover, bulge colour and disc colour
appear to be correlated pointing toward the presence of secular
evolution processes \citep{PB96,WGF97,Peletal99a,GdA01}. Based on
these indications, \citet{KK04} conclude that stellar populations are
at least consistent with the expectation that the latest type galaxies
must have pseudobulges.

The aim of this paper is to look into this in more detail, and to
search for fingerprints of recent star formation in bulges by deriving
average ages and abundance ratios of bulges along the Hubble sequence.
We the use of the \aFe\ ratio as a measure for Type~Ia supernova
enrichment, and hence late star formation.  We study the sample of
\citet[][hereafter PS02]{PS02} comprising 32 spheroids in a relatively
large range of Hubble types from E to Sbc.  By means of abundance
ratio-sensitive stellar population modelling \citep*{TMB03a,TMK04}, we
derive luminosity weighted ages, metallicities, and \aFe\ ratios of
the central stellar populations (inner $\sim 250\;$pc) from a
combination of metal indices (\Mgb, Fe5270, Fe5335) and Balmer line
indices (\HdA, \HgA, \Hb). The resulting stellar population parameters
are compared with the results obtained by PS02, and then analysed with
two-component models with the aim to quantify the possible
contribution of recent star formation on the basis of a generally old
population.

An additional constraint will be set by the direct comparison with
recent findings on early-type galaxies \citep{Thoetal05} under the
premise that a deviation of bulge properties from those of early-type
galaxies may provide further hints on the possible presence of secular
evolution and pseudobulge components.

The paper is organised as follows. After a brief summary of previous
work in the literature on the stellar populations in bulges
(Section~\ref{sec:previous}), we will first present stellar population
parameters ages, metallicities, and \aFe\ ratios derived from Mg-, Fe-
and the three Balmer line indices and compare them with the results of
PS02 (Section~\ref{sec:parameters}). In Sections~\ref{sec:comparison}
and~\ref{sec:sfhs} we will confront bulges with early-type galaxies
and derive star formation histories to quantify the amount and epoch
of possible rejuvenation events. The results are discussed in
Sections~\ref{sec:discussion} and~\ref{sec:conclusions}.

\section{Previous and current work}
\label{sec:previous}

\subsection{Colours}
In the last decade, a number of papers have analysed colours of spiral
bulges. Studying optical and near-infrared (near-IR) colour maps,
\citet{BP94} and \citet{Peletal99a} find that bulges of early-types
are as old as elliptical galaxies, while smaller bulges in later-type
spirals do show bluer colours and hence evidence for more recent star
formation \citep[see also][]{MacArtetal04}. In particular the steeper
colour gradients in the latter point to the fingerprint of the galaxy
disc and secular evolution of the bulge. This possibility is further
supported by the finding that fainter bulges have exponential profiles
\citep{CSM98,Baletal03}, are more elongated \citep{FP03}, and are more
deeply embedded in their host disk than earlier type bulges
\citep*{MACH03}. These properties clearly distinguish late-type bulges
from both their counterparts in early-type spirals and elliptical
galaxies, pointing to the presence of secular evolution in later-type
systems.

\subsection{Kinematics}
The Fundamental Plane provides a similar picture. Even though no
significant difference is found between bulge rotational properties
and ellipticals with similar absolute magnitudes
\citep{KI82,Davetal83} and ellipticals and bulges form a common major
sequence in the $\kappa$-space, the latter are slightly offset
indicating lower M/L ratios \citep*{BBF92}. This offset seems to
increase going to later types \citep*{FPB02}. These results imply that
bulges and ellipticals must have had a common (or similar) formation
epoch, while later type systems are increasingly affected by secular
evolution. \citet{CB04} analyse boxy bulges in disc galaxies with
types S0 to Sbc, and find a large fraction of bar-like structures
pointing toward a formation processes involving disc material. Hence,
both studies of colours and kinematics of bulges clearly support
secular evolution models.

\subsection{Absorption line indices - past}
The situation becomes much more confused when absorption line index
data are taken into consideration. Earlier work analysing Mg- and
Fe-indices concludes that bulges are \aFe\ enhanced just like
elliptical galaxies \citep{BP95,FFI96,IFC96,JMA96,Casetal96}. Also
recent results of the Ca triplet absorption around $8600\;$\AA\
suggest no, or very small, differences between ellipticals and bulges:
\citet{Faletal03} find that the latter seem to fit nicely into the
CaT$^*$-$\sigma$ anti-correlation established by giant and dwarf
elliptical galaxies \citep{Sagetal02,Cenetal03,Micetal03}.  At face
value, these results suggest little difference between bulges and
ellipticals, which works against the idea of secular evolution in
terms of disc-triggered star formation. Still, details do reveal some
interesting peculiarities.  \citet{BP95} find that bulges of S0s show
a large spread in Mg/Fe ratio, while the discs are mostly consistent
with solar element ratios. Most interestingly, those bulges which have
low Mg/Fe ratios also have younger ages. It seems this supports
secular evolution. Note, however, that those results refer to
lenticular galaxies, while secular evolution is expected to be
increasingly important for bulges of later-type spirals \citep{KK04}.

\subsection{Absorption line indices - present}
Clearly, new data sets are needed. Three major projects, two of them
yet to be completed, aiming at the study of bulges in spiral galaxies
can be found in the current literature.

\citet*{TDW99} plan to analyse a relatively large sample of 91 face-on
spirals comprising a huge range of galaxy types from S0 to as late as
Sdm. The sample includes both barred and unbarred spirals. Long-slit
observations along both the major and the minor axes are carried out
at the Las Campanas Observatory (6.5m) giving relatively high
resolution ($2\;$\AA) in the $4000-5200\;$\AA\ region. First results
derived from 10 objects indicate that massive bulges of earlier type
spirals are old and metal-rich like elliptical galaxies, while the
smaller bulges of later types appear younger and more metal-poor
\citep{TDW99}. These results are in line with the colour studies
basically supporting secular evolution, but the role of contamination
from disc light still needs to be assessed properly.

\citet*{GGJ99} and \citet*{JGG02} have collected a sample of 28
spirals of types from S0 to Sc. Covering a similar wavelength range as
the project above, exposure times of about 4h per object at a 4m-class
telescope allow to study gradients of all important absorption
features out to at least $1\ R_{\rm e}$. In contrast to \citet{TDW99},
these authors select edge-on spirals rather than face-on. The idea is
that placing the slit along the minor axis of an edge-on spiral, one
obtains the real gradient of the bulge, only the very central pixels
being contaminated by disc light. In particular for gradient studies,
this approach should be superior over the face-on data sample in
disentangling bulge and disc components. In a first analysis, bulges
are found to have similar ages and Mg/Fe ratios like ellipticals
\citep{JGG02}. Most interestingly, the index gradients measured are
independent of the Hubble type of the galaxy. These findings provide
clear evidence against the secular evolution picture.

\begin{figure*}
\begin{center}
\includegraphics[width=0.4\linewidth]{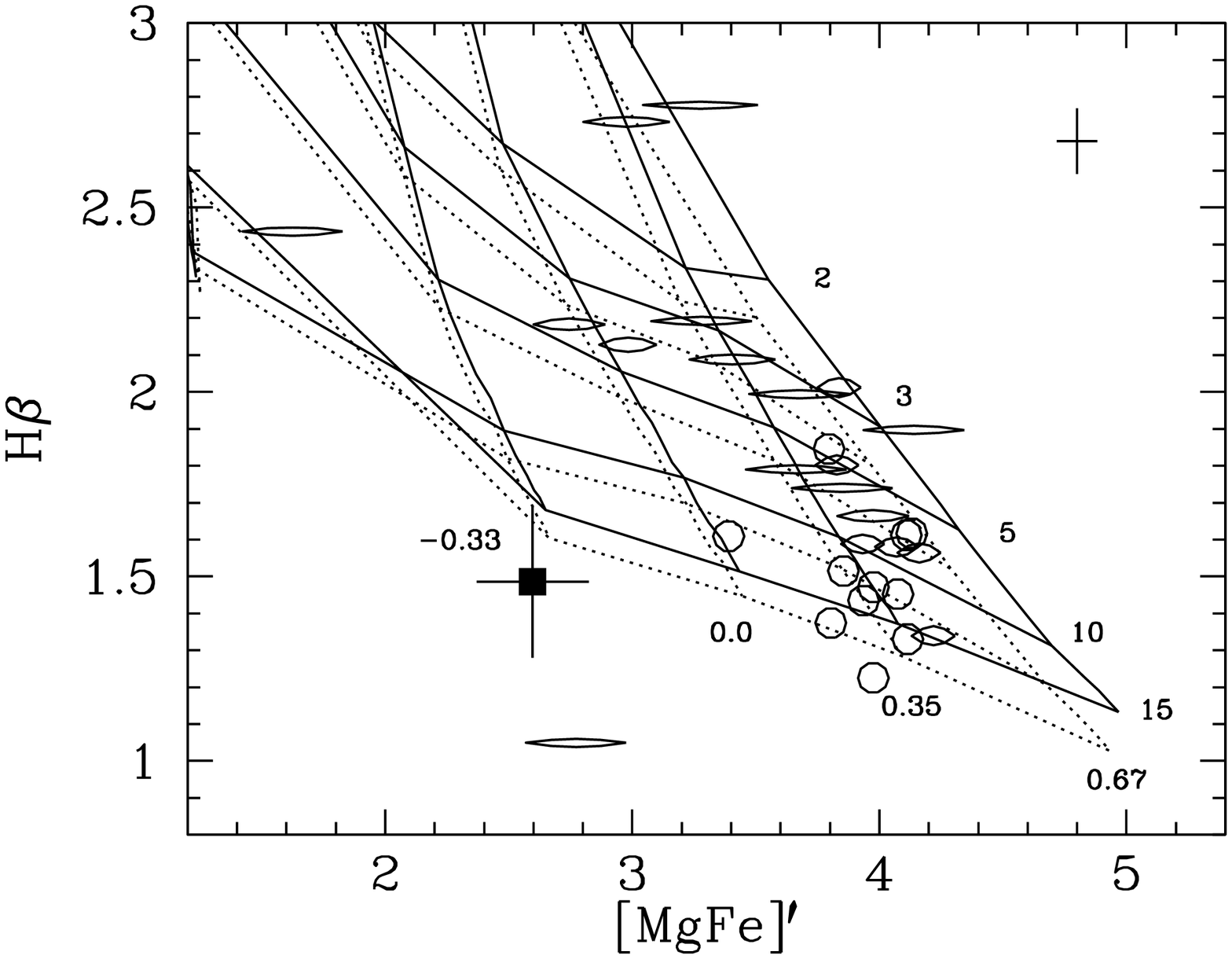}\hspace{3mm}
\includegraphics[width=0.4\linewidth]{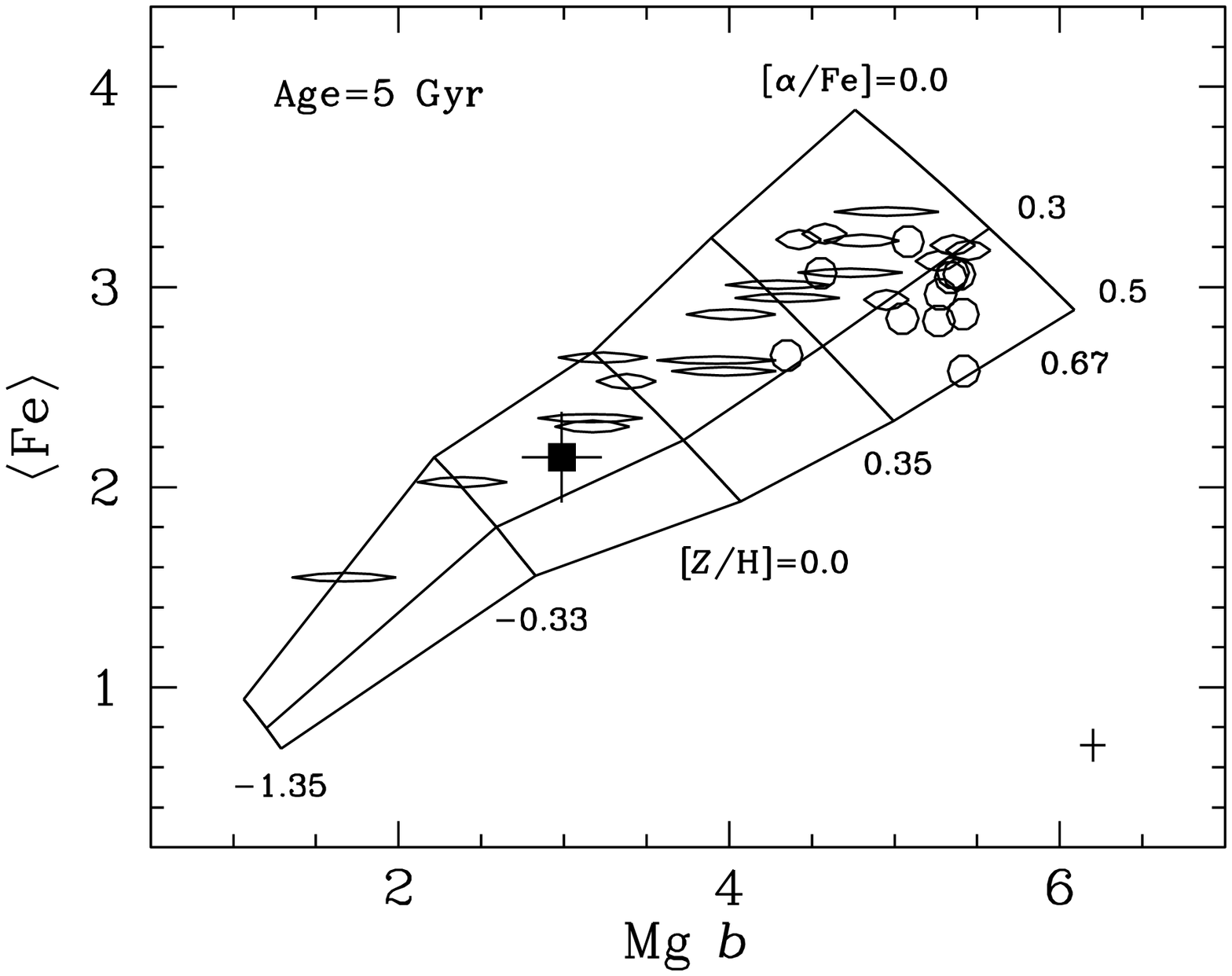}\\[3mm]
\includegraphics[width=0.4\linewidth]{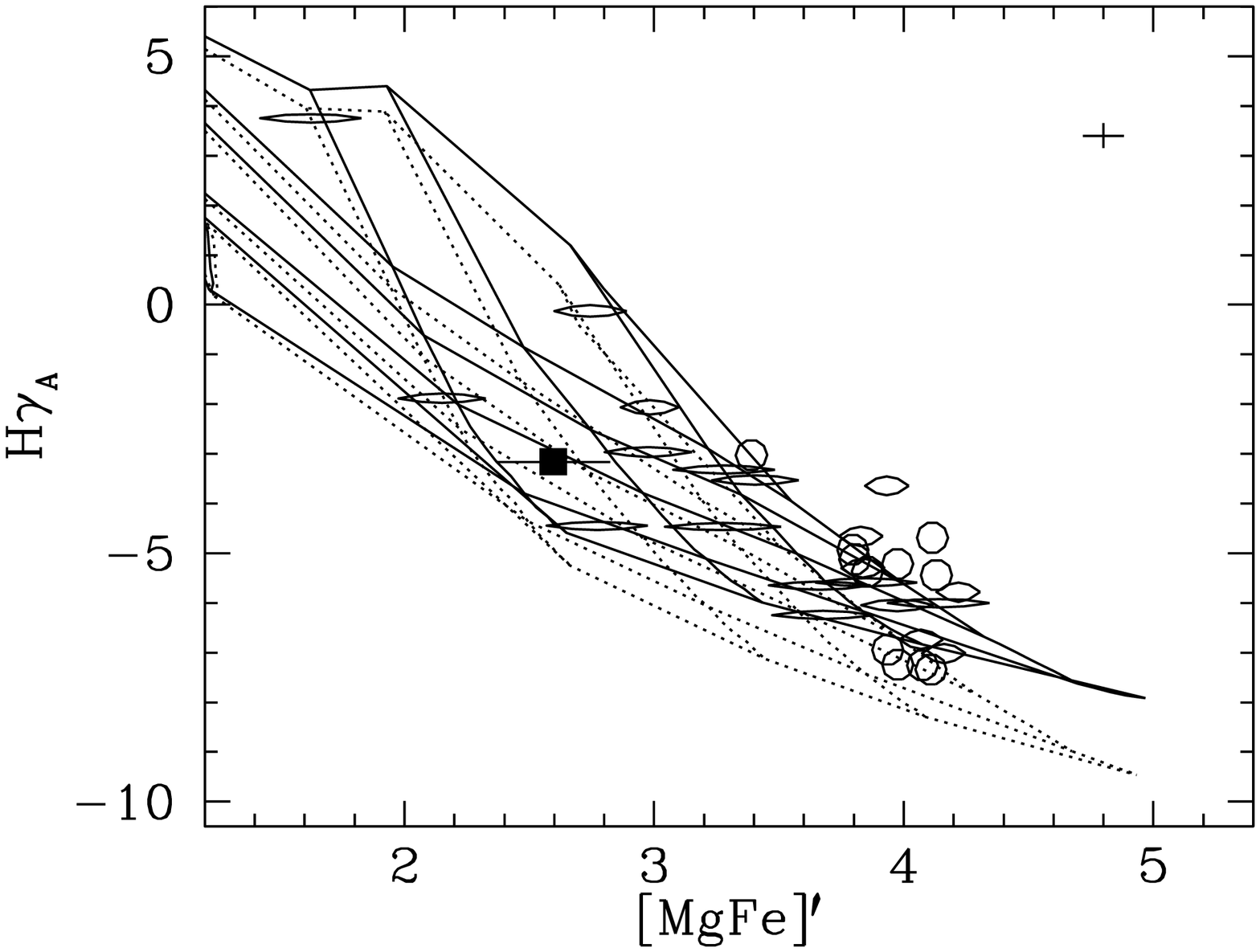}\hspace{3mm}
\includegraphics[width=0.4\linewidth]{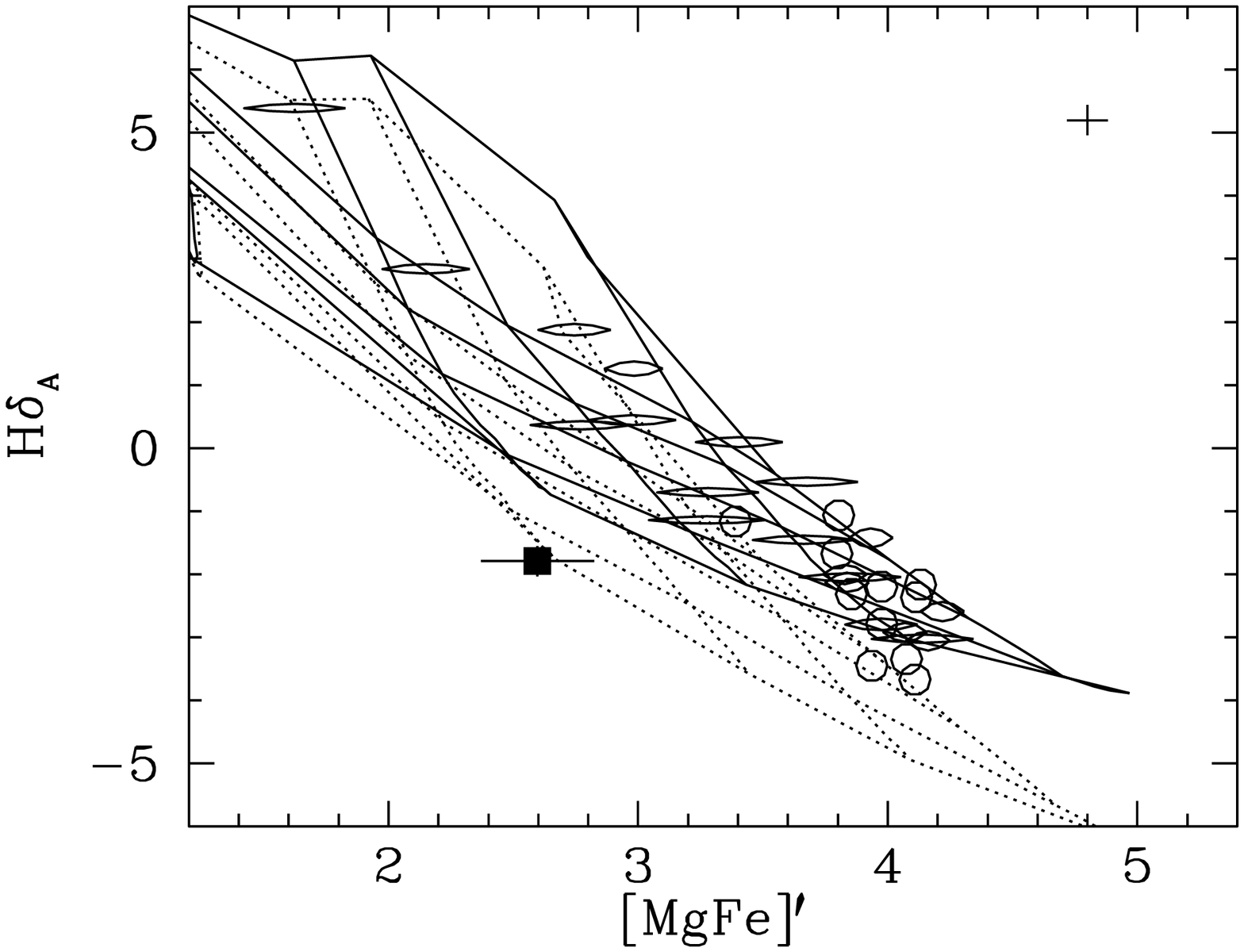}
\end{center}
\caption{Lick absorption line indices (inner $\sim 250\;$pc) of
early-type galaxies (circles) and spiral bulges (ellipses with
ellipticity increasing for the later types) from \citet{PS02}.  The
filled square is the integrated light of the Milky Way Bulge from
\citet{Puzetal02}. Stellar population models \citep{TMB03a,TMK04} for
various ages, metallicities, and \aFe\ ratios (see labels) are over
plotted. In the left-hand and the right-hand bottom panels dotted and
solid lines are models with solar and enhanced ($[\aFe]=0.3$)
abundance ratios, respectively. In the right-hand top panel age is
fixed to 5$\;$Gyr.
\label{fig:indices}
}
\end{figure*}
Finally, PS02 present a sample of 16 bulges in spirals with types Sa
to Sbc, 6 lenticular and 11 elliptical galaxies. Like in the previous
study, highly inclined targets are chosen. Slits are placed along the
minor axes trying to avoid dust lanes, the wavelength range covered is
similar to that of the studies above. The relatively modest exposure
times ($\la 1\;$h) at a 4m-class telescope make this data sample less
suitable for gradient studies. The main strength of this sample is the
coverage of galaxy types, as the inclusion of lenticular and
elliptical galaxies makes a very direct comparison with bulges
possible. PS02 report both ages and \aFe\ ratios of bulges to be lower
than those in elliptical galaxies. A metallicity-mass relationship
seems to be defined only by the bulges, and sharp differences between
early and late-type bulges are apparent. These might be indication for
secular evolution, even though this point is not discussed in the
paper.

\section{Stellar population parameters}
\label{sec:parameters}
Fig.~\ref{fig:indices} presents the Lick absorption line indices \HdA,
\HgA, \Hb, \Mgb, and \Fe\ citep{Woretal94,Traetal98} of the PS02
sample in comparison with the stellar population models of
\citet{TMB03a,TMK04} for various ages, metallicities, and \aFe\ ratios
as indicated by the labels. The most important characteristic of these
models is the inclusion of element abundance ratio effects using
metallicity-dependent index response function calculated on
high-resolution model atmospheres by \citet*{KMT05}.  Open circles are
early-type galaxies (ellipticals and S0s), bulges are the ellipses
with ellipticity increasing for the later types. The filled square is
the integrated light of the Milky Way Bulge \citep{Puzetal02}.  These
diagrams can already be used for first tentative conclusions on
$V$-luminosity weighted ages and element abundances. The sample quite
clearly forms an age sequence, bulges being both younger and less
\aFe\ enriched than ellipticals and S0s as already concluded by PS02
and most of the studies summarised in Section~\ref{sec:previous}.

\subsection{Balmer line indices and element abundance ratios}
We want to draw the attention to another, more technical, but
extremely important issue: the consistency of age estimates from the
various Balmer absorption indices \HdA, \HgA, and \Hb. It is shown in
\citet{TMK04} that the higher-order Balmer line indices \citep{WO97}
become very sensitive to \aFe\ element ratios at metallicities above
solar. More specifically, the index strength significantly increases
with increasing \aFe\ ratio. The origin for this behaviour is the
large number of Fe-lines lines in the pseudo-continua of the index
definitions. A decrease of Fe abundance leads to a higher
pseudo-continuum, and hence an increased index strength
\citep{TMK04,KMT05}. As the \aFe\ enhanced models are characterised by
a decrease in Fe abundance, this results in a positive correlation of
index strength and \aFe\ ratio. It is demonstrated that with the
inclusion of this effect, consistent ages are derived for the
\citet{KD98} early-type galaxy sample from \Hb\ and \HgA/\HgF.  Note
that, the bluer the wavelength range considered, the more Fe lines
perturb the Balmer index, and a larger abundance ratio effect is
predicted \citep{TMK04}.

This can be appreciated in Fig.~\ref{fig:indices}. The dotted and
solid lines are models with solar and enhanced ($[\aFe]=0.3$)
abundance ratios, respectively. While the size of the effect is
comparable to the measurement errors and negligible for \Hb, \HdA\
clearly is most affected. Most importantly, the solar-scaled models
(dotted lines) yield highly inconsistent age estimates from the three
Balmer lines, a problem which disappears when the abundance ratio
effect is taken into account (solid lines). A more quantitative
comparison is presented in the following section.

\subsection{Results from the various Balmer indices and PS02}
\begin{figure*}
\includegraphics[width=0.7\linewidth]{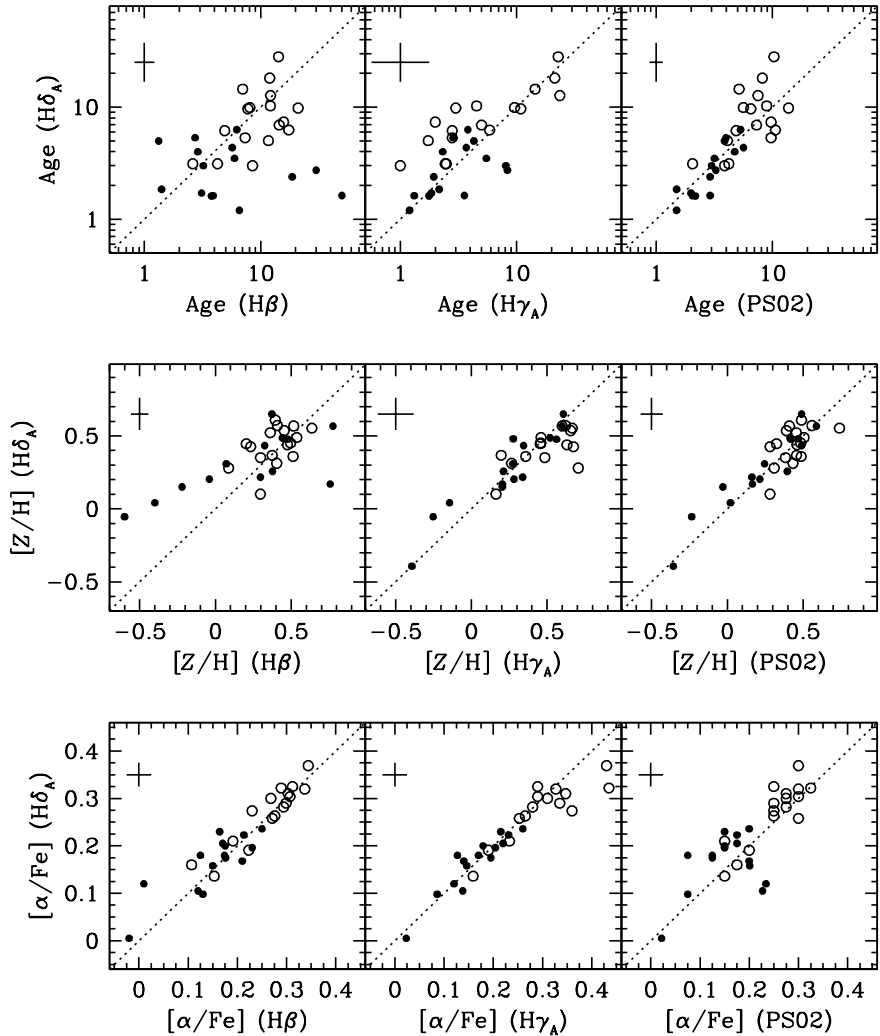}
\caption{Comparison of the stellar population parameters age,
metallicity, and abundance ratios derived from the \HdA\ as functions
of those when \Hb\ or \HgA\ are used instead. The right-hand panels
show the comparison with the results from \citet{PS02}. Open circles
are early-type galaxies, filled symbols are bulges.
\label{fig:comparison}
}
\end{figure*}
We derive the luminosity weighted stellar population parameters age,
metallicity, and \aFe\ ratio from the metallic indices \Mgb, Fe5270,
Fe5335 plus a Balmer line index (\HdA, \HgA, or \Hb) adopting the
iterative procedure described in detail in \citet{Thoetal05}. The aim
of this section is to check consistency between the three Balmer
indices and to confront the results with PS02. It should be noted
beforehand that PS02 have corrected all three Balmer line indices for
emission line filling adopting relationships between the Balmer and
the [O{\sc iii}] emission lines.  Fig.~\ref{fig:comparison} relates
age (top panels), metallicity \ZH\ (middle panels), and \aFe\ (bottom
panels) derived using the higher-order Balmer line index \HdA\ with
the ones obtained with \Hb\ and \HgA. The comparison with the results
of PS02 is shown in the right-hand panels. Bulges are now plotted as
filled circles without indicating the Hubble type to enhance the
visibility of the plot. The dotted lines indicate identity.

\subsubsection{\HdA\ -- \Hb}
Ages agree fairly well, but the scatter is large. For three bulges
(NGC~4157, NGC~4217, NGC~4312) \Hb\ indicates significantly older
ages, all three well above the age of the universe. Interestingly,
exactly those objects show H$_{\beta}$ emission greater than expected
from [O{\sc iii}] (PS02). Therefore their ages must clearly be
overestimated, a problem by which the higher-order Balmer line \HdA\
is much less affected. In these three cases \Hb\ yields metallicities
that are too low in line with the overestimated ages. The cases in
which H$_{\beta}$ emission appears lower than expected from [O{\sc
iii}] (NGC~3254, NGC~3769, NGC~4313) do not exhibit any systematic
deviation between \Hb- and \HdA-ages, but display a particularly large
scatter. Otherwise, metallicities agree well. The \aFe\ is quite
insensitive to these problems, the agreement is very good.

\subsubsection{\HdA\ -- \HgA}
For the bulges, ages agree much better between \HgA\ and \HdA. In
particular the emission line problems discussed above become much less
severe, as expected, because \Hg\ is significantly less affected by
emission than \Hb\ (see PS02 and references therein). NGC~4157,
NGC~4217, NGC~4312 still show slightly older ages from \HgA, but the
deviation approaches the general scatter in the relationship.  For
about half of the early-type galaxies the ages are systematically
underestimated by \HgA\ with respect to \HdA. From
Fig.~\ref{fig:indices} (bottom panels) it can be seen that these are
the objects that scatter above the model grid in the \MgFep-\HgA\
plane. However, perfectly consistent ages are derived for the other
half. As there are no systematic differences in the stellar population
parameters between the two groups, we interpret this deviation as
observational scatter and a probable slight underestimation of
observational errors. The younger ages of the 'deviating group' go
along with slightly lower metallicities and significantly higher \aFe\
ratios. The latter comes from the high sensitivity of \HgA\ to the
abundance ratio. For the majority of the sample (including all
bulges!), the estimates of both \ZH\ and \aFe\ from the two
higher-order Balmer lines agree very well with each other.

\subsubsection{\HdA\ -- \ PS02}
Finally, the right-hand panels in Fig.~\ref{fig:comparison} show the
comparison with the results of PS02. It is worth recalling here that
PS02 do not use a specific Balmer line, but perform a minimum $\chi^2$
fit to all 25 Lick indices. This approach can be considered fully
complementary to ours. While we use only the few line indices that we
understand and model very well, PS02 average out the ignorance of the
detail by using the maximum possible information available.  Hence it
comes somewhat as a surprise that the ages derived here with \HdA,
\Mgb, Fe5270, and Fe5335 agree so well with those of PS02. This result
is highly reassuring and suggests that these two rather orthogonal
methods yield correct results. This conclusion gets further support
from the study of \citet*{PFB04}, who find good consistency between
the ages of globular clusters derived from the Balmer line indices
separately and their minimum $\chi^2$ method when using the
\citet{TMB03a} models (see their Fig.~5).  There is some scatter in
the ages of the early-type galaxies, which most likely is scatter in
our age derivation caused by the combination of observational errors
in \HdA\ and the narrowness of the model grid at high metallicities
(see Fig.~\ref{fig:indices}). It should be emphasized that we obtain
almost identical ages for all the objects with suspicious emission
line patterns (see above), which had been responsible for deviations
in \Hb- and \HgA-ages.

Likewise, metallicities agree.  The \aFe\ ratios are in reasonable
agreement, and most importantly no systematic discrepancies are
detected. This does not come as a surprise as in both the
\citet{TMB03a,TMK04} models and the work of PS02 the incorporation of
abundance ratios is based on a similar method \citep{Traetal00a}.

Still, some scatter is present, in particular for the bulges. These
differences occur because the Thomas et al.\ models go into
significantly more detail than the method used by PS02 based on the
\citet{Vazdekis99} stellar population models. A major difference
concerns the distinction of the various evolutionary phases for the
inclusion of the index responses to abundance ratios variations from
\citet{TB95}. These are given for dwarfs, turnoff stars, and
giants. Thomas et al.\ apply these in the various phases separately
before the final stellar population is synthesised. PS02, instead,
apply the correction to the synthesised model weighting the index
responses with a constant giant-turnoff-dwarf ratio of 53-44-3. The
latter reflects the contributions to the flux continuum from the
various evolutionary phases. Hence, this approximation does not
account for the temperature and gravity sensitivity of the indices,
which causes significant deviations from the giant-turnoff-dwarf ratio
quoted above.  For instance, as shown in \citet{Maretal03}, in a 15
Gyr old stellar population with solar metallicity, the dwarfs do
contribute 20 per cent to the final Mg index, despite the low
continuum flux. This contribution even increases to 60 per cent in the
most metal-poor case.

\begin{figure*}
\begin{center}
\begin{minipage}{0.95\linewidth}
\includegraphics[width=0.49\linewidth]{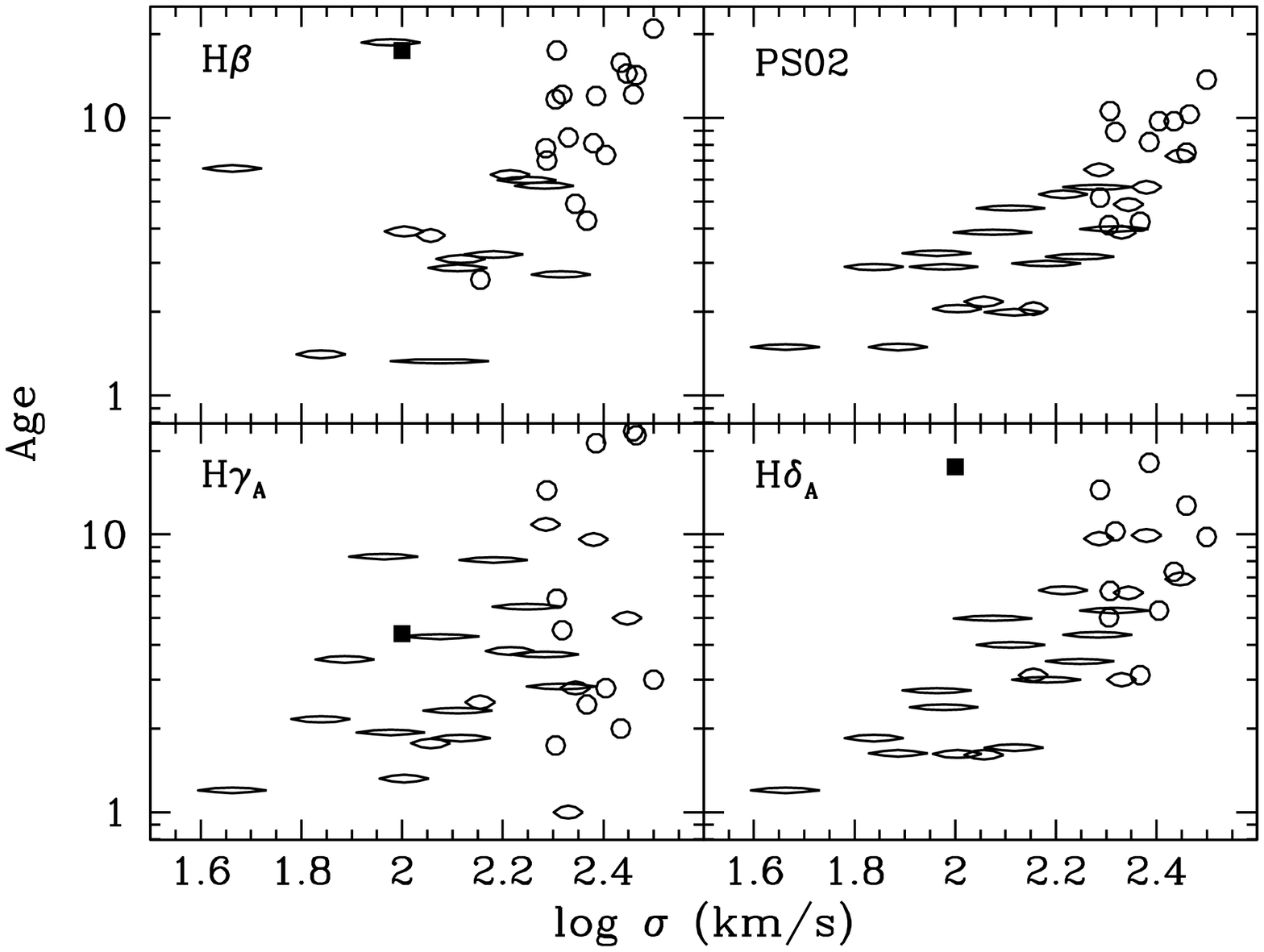}\hfill
\includegraphics[width=0.49\linewidth]{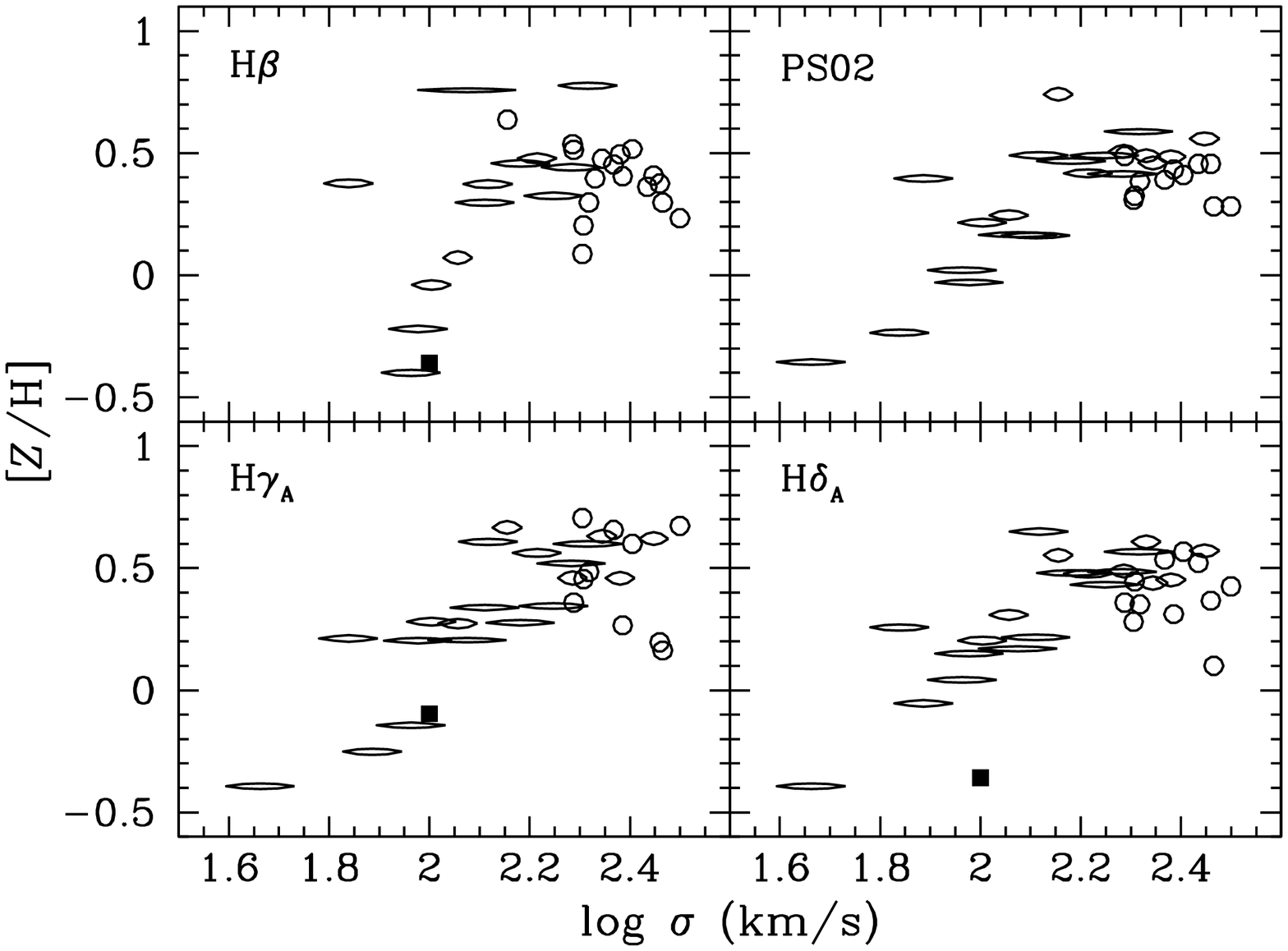}\\[2mm]
\includegraphics[width=0.49\linewidth]{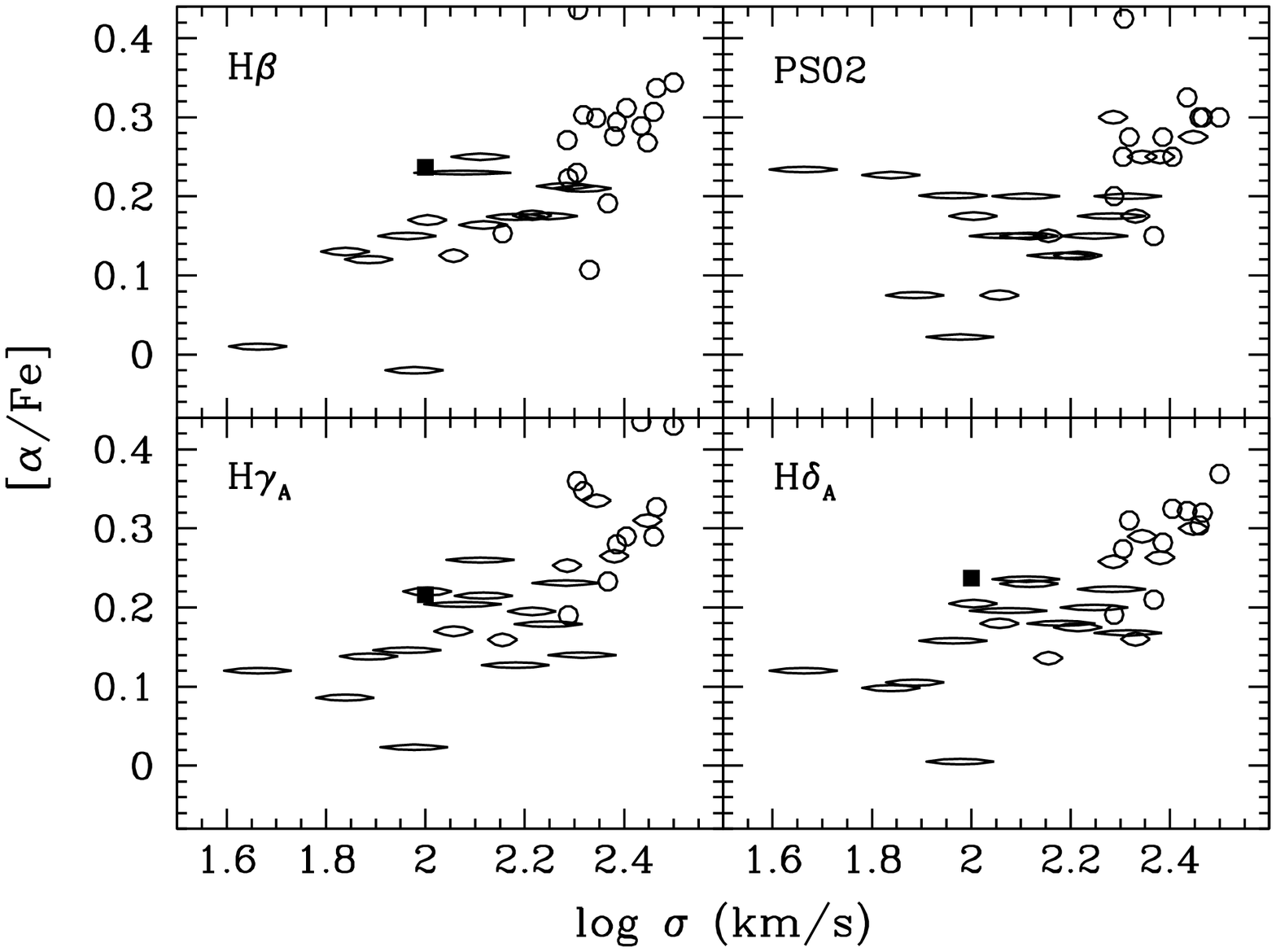}
\hfill\mbox{}
\end{minipage}
\end{center}
\caption{Comparison of the stellar population parameters age,
metallicity, and abundance ratio obtained with various Balmer line
indices (\Hb, \HgA, or \HdA\ as indicated by the labels) as age
indicators as a function of velocity dispersion. A fourth panel shows
the results of PS02. Circles are early-type galaxies, ellipses are
spiral bulges with ellipticity increasing for the later types, and the
filled square is the integrated light of the Milky Way Bulge. Errors
are indicated in Fig.~\ref{fig:comparison}.
\label{fig:sigcomp}}
\end{figure*}
A further strength of the Thomas et al.\ (updated) models is the use
of the new index response function by \citet{KMT05}, which are
calculated for the whole range of metallicities in contrast to
\citet{TB95}, who are restricted to solar. To conclude, the scatter is
most likely produced by the difference in the stellar population model
adopted, the \aFe\ ratios derived here based on the Thomas et al.\
models being more precise.

\subsubsection{Relations with velocity dispersion} 
To put the comparison discussed above into a more scientific context,
in Fig.~\ref{fig:sigcomp} we plot the stellar population parameters
resulting from \HdA, \HgA, and \Hb\ separately as functions of
velocity dispersion $\sigma$. A fourth panel shows the results of
PS02.

Both PS02 and our \HdA-ages display a very tight correlation with
$\sigma$. The other two Balmer line indices yield the same
relationship, but with significantly larger scatter. In particular
some of the bulges with old \Hb-ages fall back on the relationship
when \HdA\ is used instead. As mentioned earlier, those objects have
indeed greater H$_{\beta}$ emission than expected from the [O{\sc
iii}] line, which results in an under correction and therefore an
underestimation of the real Balmer line strength. In the other cases
the emission line correction has obviously been successful but has
significantly increased the systematic error and hence the scatter
about the relation. The same holds for \HgA, as this index turned out
to be as sensitive to the emission line problem as \Hb, because of a
reduction in the continuum level (see PS02). \HdA\ is clearly to be
preferred as age indicator, once the \aFe\ effect is taken into
account in the models. The extraordinary good consistency with the
ages of PS02 further supports this conclusion.

Total metallicities derived from \HgA, \HdA, and by PS02 agree well
and show a clear correlation with $\sigma$. Only the metallicities
obtained through \Hb\ seem not to be accurate enough to display this
relationship. In case of the \aFe\ ratios the situation is different.
All three Balmer line indices yield clear \aFe-$\sigma$ relationships
with comparable scatter, which is simply because the \aFe\ ratio
determination does not crucially depend on the derived age.  The
results of PS02 are in overall reasonable agreement with this.
Except, PS02 find \aFe\ ratios for NGC~3769 and NGC~4312 that are more
than 0.1~dex higher than ours and seem relatively high given the low
velocity dispersions of these objects.  This discrepancy at the
low-$\sigma$ end dilutes the otherwise quite clear correlation between
\aFe\ ratio and $\sigma$ in the PS02 results.

\begin{figure*}
\includegraphics[width=0.85\linewidth]{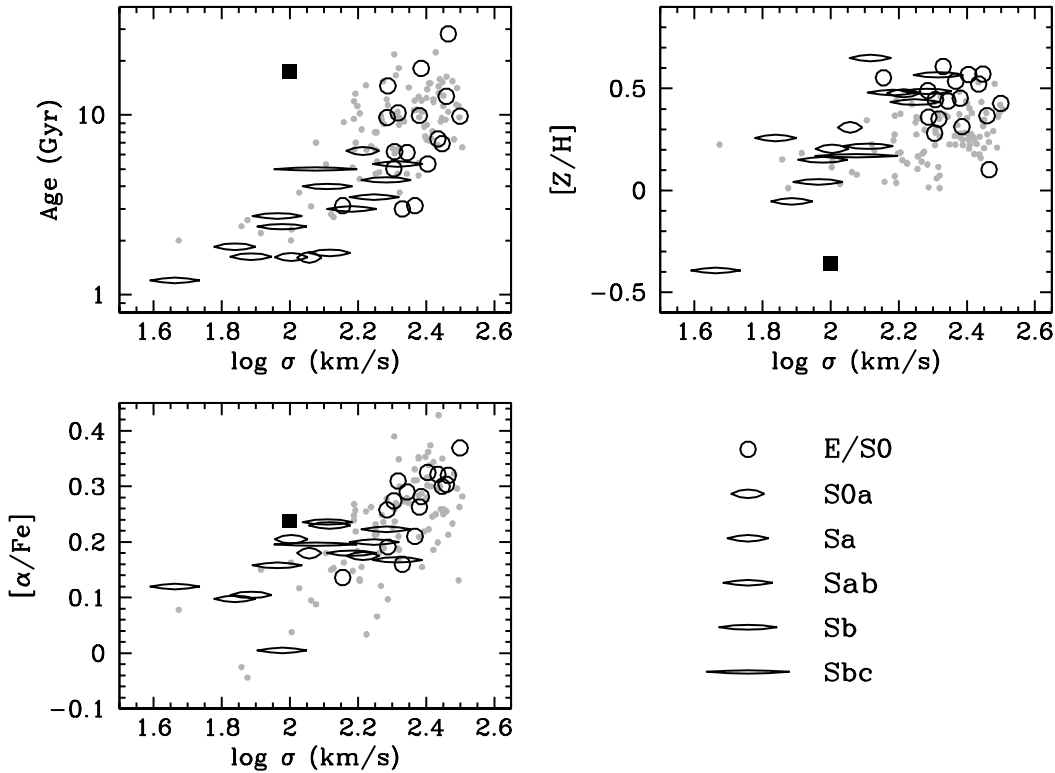}
\caption{Stellar population parameters vs.\ velocity dispersion. Open
circles are early-type galaxies from this work, ellipses are spiral
bulges with ellipticity increasing for the later types (see labels in
the right-hand bottom panel), and the filled square is the integrated
light of the Milky Way Bulge.  Small grey filled circles are
early-type galaxies from \citet{Thoetal05}. Central stellar
populations are shown.  Errors are indicated in
Fig.~\ref{fig:comparison}.
\label{fig:Ecomp}}
\end{figure*}
\subsubsection{The Bulge of the Milky Way}
The filled square in Fig.~\ref{fig:sigcomp} shows the result for the
integrated light of the Milky Way Bulge \citep[line index data
from][]{Puzetal02}. As shown in \citet{Maretal03}, the Bulge has
relatively low Balmer line indices indicating a high luminosity
weighted age and a low total metallicity (see Fig.~\ref{fig:indices}),
which significantly deviate from the age- and \ZH-$\sigma$ relations
found for the other bulges. Reasonable consistency is obtained only
for the \aFe\ ratio, as it depends only little on the Balmer line
index.  This result is in good agreement with the old ages and in
reasonable agreement with the mean metallicity obtained by
\citet{Zocetal03} from a near-IR colour-magnitude diagram of the
Bulge. However, the various Balmer indices do not provide consistent
results.  \citet{Puzetal02} measure a somewhat stronger \HgA, which
yields younger ages (and hence higher metallicities) that would be in
much better agreement with the rest of the bulges
(Fig.~\ref{fig:sigcomp}).  Most interestingly, this young
'\HgA-solution' is consistent with the parameters derived by
\citet{PFB04} from the same data based on the $\chi^2$ technique.

\section{Bulges and ellipticals in comparison}
\label{sec:comparison}
The sample of PS02 contains both bulges and early-type galaxies. A
direct comparison between the two subsets clearly suggests that bulges
are younger, less metal-rich and less \aFe\ enhanced. However, the
bulges in the sample have systematically lower central velocity
dispersions $\sigma$, and very clear correlations of these three
parameters with $\sigma$ are evident. As all three parameters are
known to correlate for early-type galaxy samples \citep[e.g.][and
references therein]{FS00,Traetal00b,Neletal05,Thoetal05,Beretal05},
and it is not clear, whether the difference in the stellar population
properties is simply the result of this relationship or whether it
hints to different formation scenarios and disc influence in spiral
bulges.  A meaningful comparison as a function of Hubble type must
certainly be carried out at a given $\sigma$.

In the following, we therefore confront the present results with the
data of \citet{Thoetal05}, in which stellar population parameters for
a sample of early-type galaxies extending to relatively low velocity
dispersion of around 100$\;$km/s are derived.  In Fig.~\ref{fig:Ecomp}
we plot age, total metallicity, and \aFe\ ratio of both samples as
functions of velocity dispersion. For the PS02 sample (open symbols),
the results obtained with the higher-order Balmer index \HdA\ (see
previous section) are adopted, while \Hb\ is used in \citet{Thoetal05}
(filled grey circles).

\begin{figure*}
\begin{center}
\includegraphics[width=0.31\linewidth]{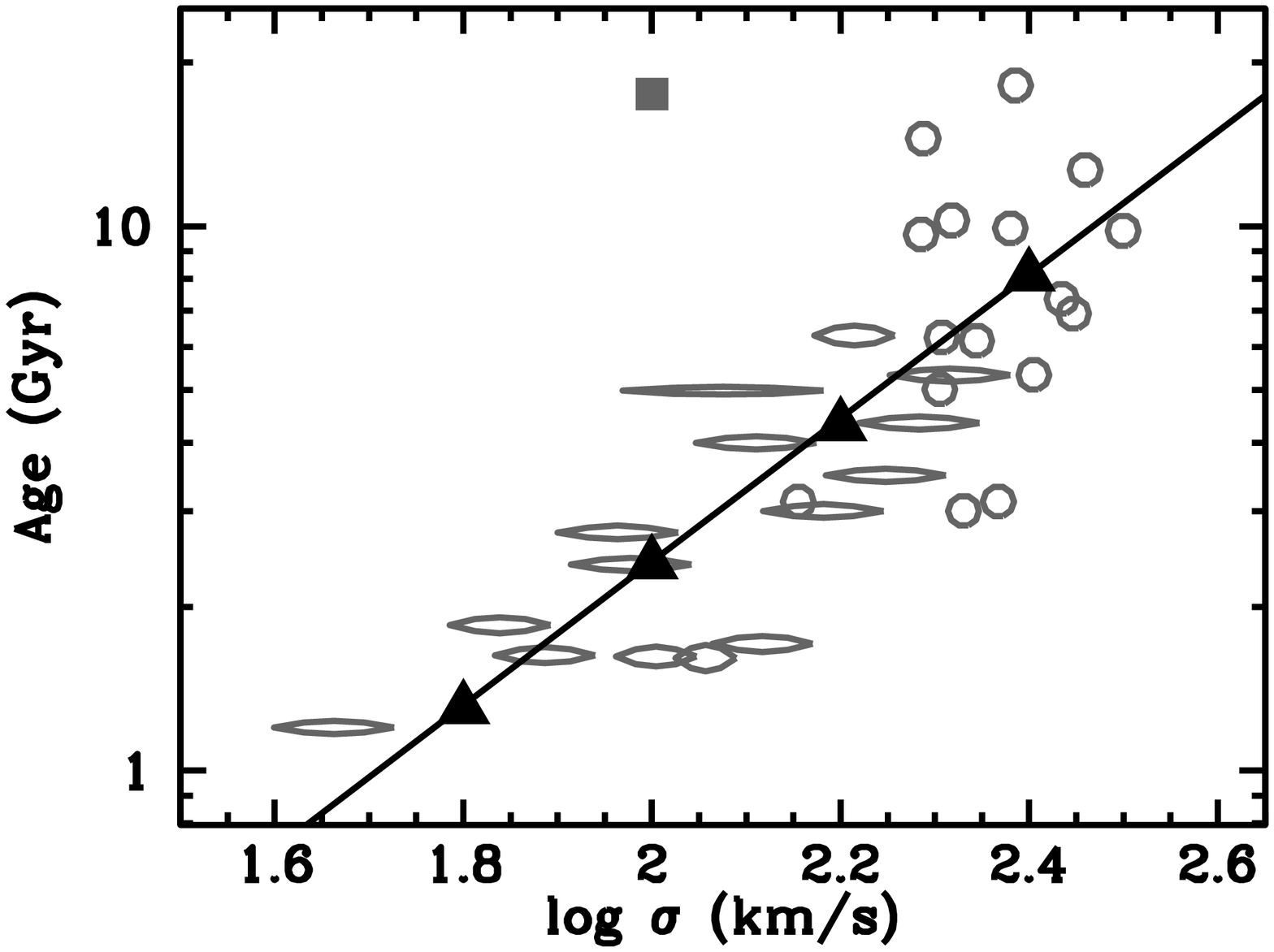}
\includegraphics[width=0.31\linewidth]{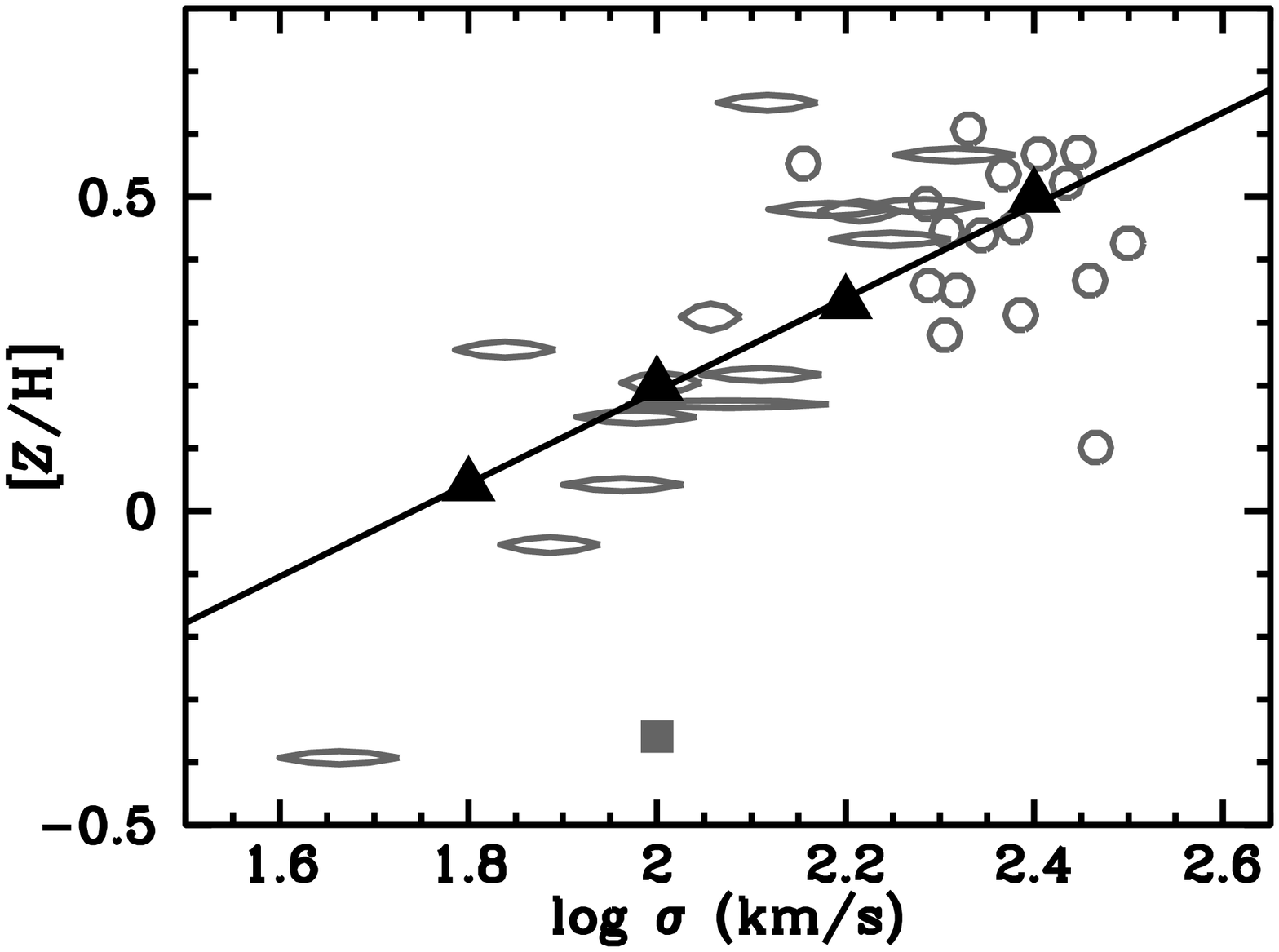}
\includegraphics[width=0.31\linewidth]{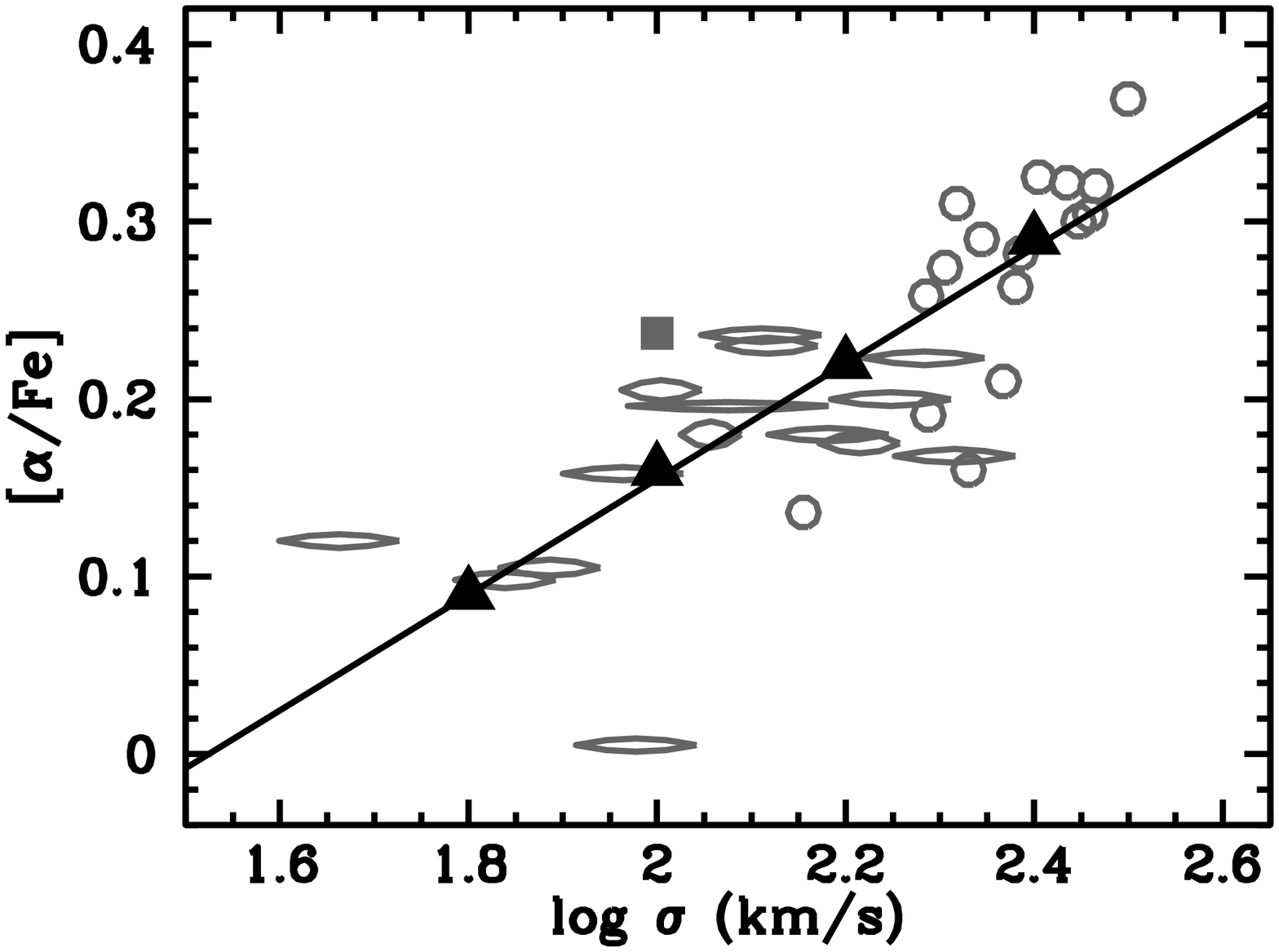}
\end{center}
\caption{Stellar population parameters as function of velocity
dispersion. Triangles are composite models considering the presence of
a young subcomponent over an underlying old population (see
Fig.~\ref{fig:sfh}). Grey symbols are the observational data points
from Fig.~\ref{fig:Ecomp}.
\label{fig:csp}
}
\end{figure*}
Before confronting the PS02 bulges with the low-mass early-type
galaxies of \citet{Thoetal05}, we check consistency at high velocity
dispersion $\log\sigma>2.3$ where both samples overlap in Hubble type.
Ages and \aFe\ ratios of the PS02 early-type galaxies (open circles)
and the \citet{Thoetal05} objects (filled grey circles) indeed agree
very well. Metallicities of the PS02 objects are higher by about
0.1$\;$dex, which most likely is an aperture effect. While PS02 use a
fixed central aperture of $3.6\times 1.25\;$arcsec$^2$,
\citet{Thoetal05} consider a variable aperture that ensures the
coverage of 1/10 of the effective radius. Hence, for large objects,
PS02 sample a more centrally concentrated fraction of the stellar
population. Given the presence of a negative metallicity gradient in
early-type galaxies
\citep{DSP93,CD94,FFI95,Sagetal00,Mehetal03,Wuetal05}, this explains
the slight offset between PS02 and \citet{Thoetal05}. Such an offset
caused by aperture effects is not to be expected in age and \aFe\
ratio, because early-type galaxies are found to have no gradients in
these parameters \citep{Davetal01,Mehetal03,Wuetal05}. Ages and \aFe\
ratios of the two samples indeed agree very well.

A similarly good consistency between the samples is evident also in
the low velocity dispersion regime ($\log\sigma\sim 2$). Low-mass
early-type galaxies and bulges appear to obey the same relationship of
age and \aFe\ ratio with $\sigma$. Spiral bulges and low-mass
early-type galaxies host stellar populations with the same young
luminosity weighted ages around 2--3$\;$Gyr. In line with these, both
types of objects display low \aFe\ ratios of about 0.1$\;$dex
reinforcing the presence of recent star formation suggested by the
young ages. Metallicities agree at low $\sigma$, which is simply due
to the fact that the smaller aperture of PS02 in smaller objects
samples a larger proportion of light more consistent with
\citet{Thoetal05}. To conclude, the very prominent correlations of
age, metallicity, and \aFe\ with velocity dispersion found for spiral
bulges holds also for early-type galaxies.

\section{Star formation histories}
\label{sec:sfhs}
\begin{figure*}
\begin{center}
\includegraphics[width=0.4\linewidth]{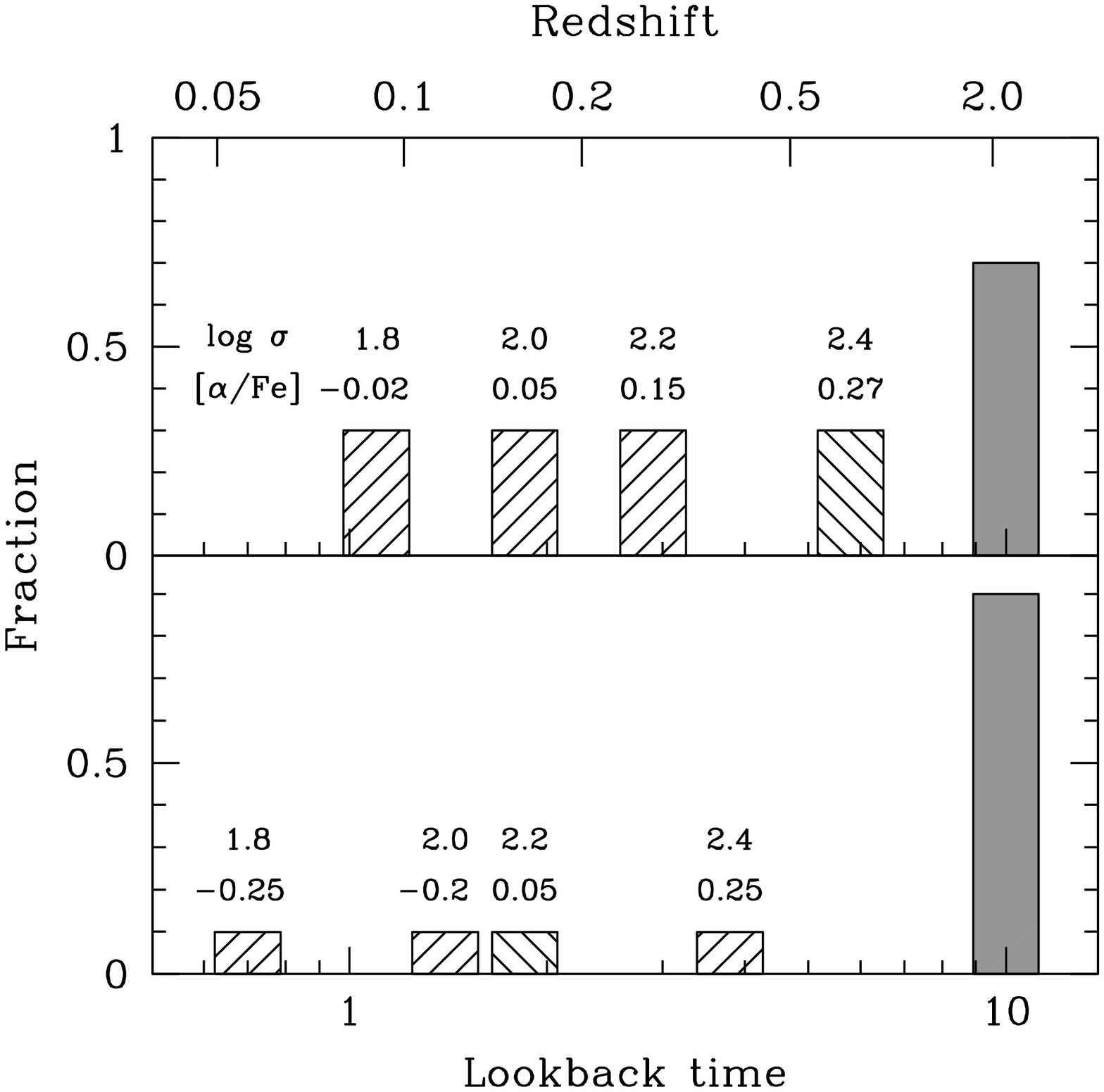}
\includegraphics[width=0.4\linewidth]{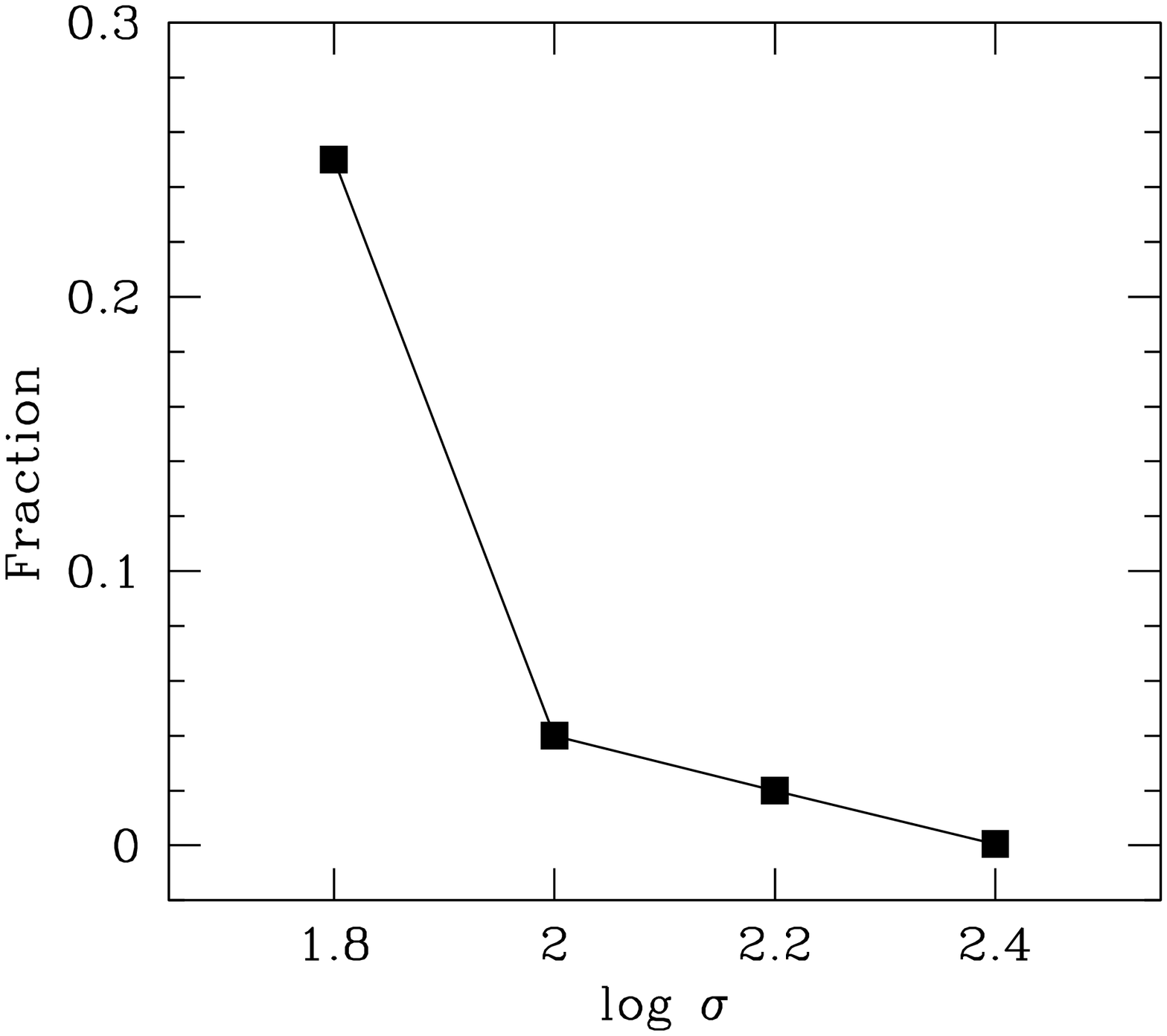}
\end{center}
\caption{Two-component star formation histories that reproduce the
observed relationships between stellar population parameters and
velocity dispersion (see triangles in Fig.~\ref{fig:csp}). An
underlying old population with a formation redshift $z_{\rm f}\sim 2$
is perturbed by a young sub-component. {\em Left-hand panel}: The mass
fraction of the secondary component (hatched rectangles) is fixed to 30
panel (top panel) and 10 per cent (bottom panel), formation redshift
is varied. Plotted are the mass fractions of the two components as
functions of look-back time and redshift ($\Omega_{rm m}=0.2,\
\Omega_{\Lambda}=0.8,\ H_0=72\;$km/s/Mpc). Labels give the associated
velocity dispersions of the various realisations and the \aFe\ ratio
of the young population. {\em Right-hand panel}: The formation
redshift is fixed to $z\sim 0.08$ (corresponding to a look-back time
of $1\;$Gyr), and the mass fraction of the young component is
varied. The plot shows this mass fraction as a function of velocity
dispersion.
\label{fig:sfh}
}
\end{figure*}
The relationships age-$\sigma$ and \ZH-$\sigma$ given in
\citet{Thoetal05} are significantly flatter than what is inferred from
Fig.~\ref{fig:Ecomp}, because they are derived for objects with
$\sigma$ well above 100$\;$km/s. As shown in \citet{Thoetal05}, the
inclusion of low-mass objects steepens the correlations for early-type
galaxies significantly \citep[see also][]{Smith05}. Indeed, in
\citet{Thoetal05} the low-mass end of the sample is best reproduced
assuming the occurrence of a minor ($\sim 10\;$ per cent) recent
($\sim 1\;$Gyr ago) episode of star formation.

\subsection{Model parameters}
Similar in spirit to \citet{DD97}, in the following we explore how the
luminosity weighted ages and the above relationships can be reproduced
by star formation histories characterised by secondary star formation
on top of a dominating, underlying old population. We assume this
latter base population to have an age of $10\;$Gyr corresponding to a
formation redshift of $z\sim 2$, and to have an abundance ratio
$[\aFe]=0.3\;$dex, reflecting a rapid formation process
\citep[see][]{Thoetal05}.  As we cannot constrain the mass
contribution from a secondary burst, we test three options: two in
which we fix the mass contribution of the recent star forming event to
10 and 30$\;$per cent, respectively.  In these two models we then vary
as a free parameter the look-back time at which the secondary burst
must have happened, in order to match the luminosity weighted age. In
the third model we fix the look-back time to $1\;$Gyr and vary,
instead, the mass fraction of the secondary population.

In general, the \aFe\ ratio of the secondary population is varied such
that the \aFe-$\sigma$ relationship of Fig.~\ref{fig:Ecomp} is
reproduced. For simplicity, metallicity is assumed to be the same for
both the base old and the secondary population, and is chosen such
that consistency with the correlation of Fig.~\ref{fig:Ecomp} is
ensured.  This approach makes the sensible assumption that velocity
dispersion (hence total mass) of the object determines the total
metallicity of the entire population. The age and element abundance
ratio, instead, are considered to be $\sigma$-dependent for the
secondary burst and universal for the base population. This prior
needs to be set, in order to avoid under-determination of the
problem. From the Balmer line index and the two metal indices we
derive age (for the first two options) or burst fraction (for the
third option) and \aFe\ ratio of the secondary population, which are
the main focus of this study. As a third parameter we obtain the
metallicity of both components, which we need for consistency reasons.

The stellar population parameters of the synthesised population are
determined as follows: first we compute line indices (\HdA, \HgA, \Hb,
\Mgb, Fe5270, Fe5335) of the two-component population. These are
obtained by summing up the fluxes in the pseudo-continua and line
windows, out of which the final index value is calculated as described
in, e.g., \citet{Maretal03}.  From these mock observables we determine
stellar population parameters. In this way we ensure that luminosity
weighted quantities are obtained.  The input parameters described
above are iteratively modified until observations as presented in
Fig.~\ref{fig:Ecomp} are reproduced.

\subsection{Consistency of the three Balmer indices}
It should be noted that the three Balmer line indices are considered
separately. As \HdA\ is located at a bluer part of the spectrum, it
responds to slightly hotter temperatures than \Hb. As a consequence, a
composite stellar population with hot sub-components must lead to
different age estimates from these two indices. The size of the
effect, however, depends strongly on the turnoff temperature (hence
the age) of the young component. In the previous section (see
Fig.~\ref{fig:sigcomp}) it is shown that consistent age estimates are
obtained.  The present simulations allow us to quantify this effect
and to test whether these consistent age estimates exclude the
presence of a hot sub-component.

The effect is maximum for a $1\;$Gyr sub-component, where the turnoff
temperature hits exactly the wavelength of the \Hd\ line. We find that
around this maximum, \HdA\ yields luminosity weighted ages that are
about $10-20\;$per cent, i.e.\ $0.05-0.08\;$dex, younger compared to
\Hb. Fig.~\ref{fig:Ecomp} shows that this difference is too small to
be detected with the present data set.  It should be emphasized that
even younger subpopulations, as recently found by \citet{Yietal05} in
some fraction of early-type galaxies on the basis of GALEX photometry,
would not affect the consistency of the ages estimated from \Hb\ and
\Hd\ as their turnoff temperatures are hotter than both \Hb\ and \Hd.

\subsection{Results}
Fig.~\ref{fig:csp} shows the stellar population parameters of the
models described above as filled triangles in comparison to the
observational data (grey symbols). We produced models for
$\log\sigma=1.8,\ 2.0,\ 2.2$ and 2.4, that -- by construction -- match
the observed correlations with velocity dispersion.

The star formation histories of these models are illustrated in
Fig.~\ref{fig:sfh}.  The left-hand panel shows the first two options,
in which the look-back time of the secondary burst is modified. The
grey filled histogram is the underlying old population with a
formation redshift of $z_{\rm f}\sim 2$. The hatched histograms show
the ages of the young sub-components (labels are velocity dispersions
and \aFe\ ratios).  If we assume the young population to contribute 10
per cent to the total mass, the ages measured here for spiral bulges
and low-mass early-type galaxies require recent star formation at
redshifts between $z\sim 0.05$ and 0.5, corresponding to look-back
times between 0.7 and $4\;$Gyr.  Assigning 30 per cent of the total
mass to the young sub-component moves these quantities only mildly to
higher look-back times as shown by the top panel. The lower $\sigma$,
the more recent is the required star formation. And the more recent
the additional star formation, the lower the \aFe\ ratio of the young
component as a larger time span allows Fe enrichment from Type Ia
supernovae (Fig.~\ref{fig:sfh}).

Alternatively, if we fix the formation redshift of the young
population, its mass fraction must increase with decreasing $\sigma$
(model 3).  This is shown by the right-hand panel of
Fig.~\ref{fig:sfh}, in which the fraction of the secondary component
is plotted as a function of $\sigma$. The contributions in mass from a
1-Gyr subcomponent increase from only 0.05 per cent for
$\log\sigma=2.4$ to as much as 25 per cent for the smallest bulges.

\section{Discussion}
\label{sec:discussion}
We analyse the central (inner $\sim 250\;$pc) stellar populations of
bulges in spiral galaxies with Hubble types Sa to Sbc by re-deriving
luminosity weighted ages, metallicities, and \aFe\ ratios of the PS02
sample.  We find that all three stellar population parameters display
very clear positive relationships with central velocity
dispersion. Our results confirm the previous finding of PS02 and, in
particular, improve upon the correlation with \aFe\ ratio.  Lower-mass
bulges have younger luminosity weighted ages and lower \aFe\ ratios,
both hinting toward the presence of extended star formation. These
results are in line with the findings of previous studies based on
colours and absorption line indices as summarised in
Section~\ref{sec:previous}. We note that the relatively young age
(5$\;$Gyr) derived for the bulge of the S0 galaxy NGC~7332
($\log\sigma\sim 2.15$) by \citet{Faletal04} fits perfectly in the
age-$\sigma$ relationship shown here. In contrast, \citet{Saretal05}
find predominantly old stellar populations in the very centres (inner
$\sim 8\;$pc) of spiral bulges, which might be caused by the fact that
the central black hole prevents star formation in its immediate
vicinity.

\subsection{Bulges -- just like low-mass ellipticals}
We show that the occurrence of recent secondary star formation
episodes involving about 10--30 per cent of the total mass at
look-back times between $\sim 0.5$ and $5\;$Gyr, corresponding to the
redshift interval $0.05\la z\la 0.5$, provide a suitable explanation
for the above relationships. The weight of the recent star formation
event, either in terms of formation redshift or mass fraction,
increases with decreasing bulge velocity dispersion. If we keep in
mind that low $\sigma$ characterises pseudobulges, this may indicate
that secondary evolution in bulges triggered by disc instabilities as
suggested by \citet{KK04} does take place, predominantly in low-mass
objects. On the other hand, the stellar population properties derived
in this paper do not depend on Hubble type, which complements the
previous findings that disc and bulge scale lengths as well as bulge
to disc ratios correlate with bulge luminosity rather than with Hubble
type \citep*{deJong96a,Couetal96,BGP05}. This, in contrast, hints to
an independent formation of the bulge rather than secular evolution.

\medskip
How can we resolve this apparent contradiction, and, more importantly,
how can we get further hints about the process at work? If the
evolution of the stellar populations in spheroids with discs around
them, i.e.\ bulges of spiral galaxies, is affected by the presence of
the disc, then spheroids without discs, i.e.\ elliptical galaxies,
should exhibit different properties. We therefore compare in this
paper the present results with the stellar population properties of
early-type galaxies \citep{Thoetal05}. It is shown in a number of
studies, that elliptical galaxies also obey a correlation between
luminosity weighted age and mass in the sense that low-mass galaxies
are affected by late star formation \citep[e.g.][and references
therein]{FS00,Traetal00b,Neletal05,Thoetal05,Beretal05,Capetal06}.

We find that bulges are generally younger than early-type galaxies,
because of their smaller masses. Bulges and early-type galaxies
exhibit the same correlation between their stellar population
properties and mass. In other words, at a given $\sigma$, we find no
difference between bulges and ellipticals, they are indistinguishable
objects as far as their basic stellar population properties are
concerned. Hence, Hubble type does not determine the stellar
populations of spheroids in a large range of Hubble type from E to at
least Sbc, which is the latest type probed in the present study. The
disc does not significantly affect the bulge's stellar
populations. Bulges, like low-mass ellipticals, are rejuvenated, but
not by secular evolution processes involving disc material. It should
be emphasized that bulges with dust lanes (like e.g., NGC~4157), that
are the most promising candidates for pseudobulges, also obey these
relationships.  Hence, secular evolution and pseudobulges can only
play a role in spiral galaxies with types later than Sbc.

\medskip
The picture of rejuvenation of bulges fits in well with the finding
that bulges are bluer than ellipticals at any given redshift out to
$z\sim 1$ \citep*{EAD01}. Our results provide a clearer insight into
this.  \citet{EAD01} find that bulges are bluer because they have both
younger ages and lower metallicities than the ellipticals in the
sample, which in turn is the simple consequence of their lower
velocity dispersions (or total masses). Indeed, in their Fig.~4
ellipticals with lower galactic concentration are as blue as the
bulges at any redshift. It is intriguing that the observed near-IR
\JH\ colours of the bulges are redder at $z\ga 0.5$, and similar to
those of ellipticals at redshifts below 0.5. The redder near-IR colour
might be connected to the TP-AGB star population present in stellar
populations with ages between 0.1 and 2$\;$Gyr \citep{Ma05}, but we
cannot think of any reasonable star formation history for which this
effect should disappear below $z\sim 0.5$. On the contrary, it should
even be more pronounced there, because the observed near-IR colour
shifts more toward rest-frame near IR, where the effect of the TP-AGB
is strongest. A more detailed investigation of this issue would be
interesting, but by far exceeds the scope of this paper.

\subsection{Caveats and open questions}
A general concern about probing the central stellar populations in the
bulge of a disc galaxy is the contamination from disc material in the
line of sight. Based on observed bulge-to-disc ratios of edge-on
spirals of type not later than Sbc \citep*{KWK00}, PS02 argue that the
disc contamination cannot exceed the 10 per cent level in
light. Because of the relatively young luminosity weighted ages
derived (consistent with PS02), the contribution of the young
subpopulation exceeds this upper limit by far. In other words, the
young ages found for bulges cannot be artifacts from projected young
disc stars. If disc contamination was artificially lowering the
derived ages, bulges would by inference be older than ellipticals at a
given $\sigma$, which would argue strongly against secular evolution.

\medskip
There is another intriguing problem. The bulge of our own galaxy is
known to be old \citep{Renzini99}. \citet{Zocetal03} analyse near-IR
colour magnitude diagrams of large, statistically meaningful, bulge
fields. They derive an age larger than $10\;$Gyr and find no trace of
any younger stellar population. Interestingly, this result is
perfectly consistent with the luminosity weighted age that we derive
in this paper (see the square in Fig.~\ref{fig:Ecomp}). Also the
sub-solar metallicity we derive is consistent with the metallicity
distribution derived by \citet{Zocetal03}. However, these values are
deviant from the general ages and metallicities of bulges. Our bulge
appears significantly older and more metal poor. This is certainly an
odd situation.

In the present work, the old age for our bulge is derived from the
\HdA\ Balmer line index, and is consistent with the age obtained from
\Hb\ (see Fig.~\ref{fig:sigcomp}). However, \HgA\ suggests a
considerably younger age of about $4-5\;$Gyr, which would then be
consistent with the age-$\sigma$ relationship of the other bulges. The
metallicity would be slightly higher, and hence also more consistent
with the derived relationship.  It is possible that the measurements
of both \HdA\ and \Hb\ are corrupted, and that the younger age implied
by \HgA\ gives the correct result. This view is supported by the
similarly young age derived by \citet{PFB04} from the same data based
on their $\chi^2$ technique. Note that the latter is quite insensitive
to uncertainties of individual absorption line indices.  This analysis
suggests that the bulge of the Milky Way is not special.

However, there is then a clear contradiction with the work of
\citet{Zocetal03}. Their conclusion is certainly robust, and the only
caveat that affects their non-detection of young stellar populations
is the correction from the contamination of foreground disc
stars. Even though this is done very carefully using a large disc
control field, an overcorrection that 'removed' young bulge stars
cannot be entirely excluded. Alternatively, the apparent conflict with
the old age found by \citet{Zocetal03} might be resolved if there was
a population gradient. The PS02 data and \citet{Puzetal02} data for
the bulge sample the central region, the inner $\sim 250\;$pc, while
the fields observed by \citet{Zocetal03} are approximately $1\;$kpc
from the centre. A rejuvenation of the centre of our bulge would
indeed fit with the fact that young stars and star clusters are found
in the very centre \citep{FMcLM99,Genetal03a}. If significant, this
would imply that bulges have positive age gradients, in contrast to
early-type galaxies for which no age gradients are detected
\citep{Davetal01,Mehetal03,Wuetal05}. Detailed studies of the
gradients in bulges may help in future to solve this issue
\citep{JGG02}.

\section{Conclusions}
\label{sec:conclusions}
The main aim of this paper is to investigate whether the evolution of
the stellar population in bulges is modified by the presence of the
disc. We seek to understand whether secular evolution, and maybe the
formation of pseudobulges, play an important role in the evolution of
spiral galaxies. Our approach is to compare the stellar population
properties in bulges with those in elliptical galaxies at a given
central velocity dispersion, hence spheroid mass.

For bulges, a suitable sample is PS02 comprising 16 bulges in spirals
with types Sa to Sbc, 6 lenticular and 11 elliptical galaxies. We
derive luminosity weighted ages, metallicities, and \aFe\ ratios from
one Balmer line index (\HdA, \HgA, or \Hb\ considered separately), and
the metallic indices \Mgb, Fe5270, and Fe5335 using the element ratio
sensitive stellar population models of \citet{TMB03a,TMK04}.  As there
is relatively little overlap in $\sigma$ between the bulges and the
early-type galaxies in the PS02 sample, we compare the results with
the sample of \citet{Thoetal05}, which contains elliptical galaxies
with velocity dispersion as low as $50\;$km/s. For both samples we
obtain very clear relationships between all three stellar population
parameters and $\sigma$.

\citet{TMK04} demonstrate that the higher-order Balmer line indices
are very sensitive to element ratios effects, and that consistent age
estimates from \Hb\ and \HgA\ are obtained only when these effects are
taken into account in the models. Here we extend this exercise to
\HdA, as the PS02 sample includes also this index.  We obtain very
consistent estimates of ages, metallicities, and \aFe\ ratios from the
three Balmer line indices. Emission line filling plays a critical
role, impacting crucially on the scatter of the derived age-$\sigma$
relation. The latter is smallest for \HdA. Importantly, the ages and
metallicities derived here using \HdA\ are extraordinarily consistent
with those given by PS02.  Note that PS02 do not use a specific Balmer
line, but perform a minimum $\chi^2$ fit to all 25 Lick indices, an
approach fully complementary to ours. While we use only the few line
indices that we understand and model very well, PS02 average out the
ignorance of the detail by using the maximum possible information
available.  The excellent consistency found here is reassuring and
suggests that these two rather orthogonal methods yield correct
results. It should be emphasized that the $\chi^2$ method is not
dominated by the particular line indices used in the present study.

In agreement with other studies, we find that bulges have relatively
low luminosity weighted ages, the lowest age derived being
$1.3\;$Gyr. Hence bulges are not overall old, but are actually
rejuvenated systems. Interestingly, there is evidence that the bulge
of the Milky Way also fits into this picture.  We find clear
correlations of all three parameters luminosity weighted age, total
metallicity, and \aFe\ ratio with central velocity dispersion, the
smallest bulges being the youngest with the lowest \aFe\ ratios owing
to late Fe enrichment from Type Ia supernovae.  We construct composite
models in which a young subcomponent is superimposed over an
underlying old population, in order to constrain the epoch and mass
fraction of the rejuvenation event. We show that the smallest bulges
must have experienced significant star formation events involving
$10-30\;$ per cent of their total mass in the past
$1-2\;$Gyr. Curiously, these results are not consistent with the age
estimates for the Bulge of the Milky Way in the literature, which
appears to have an overall old stellar population and no traces of
recent star formation. We discuss new evidence that at least the
central $\sim 500\;$pc of the Milky Way bulge contains a significant
fraction of young stellar populations.

The comparison with the \citet{Thoetal05} sample reveals that the
above relationships with $\sigma$ coincide perfectly with those of
early-type galaxies.  In other words, bulges are typically younger,
metal-poorer and less \aFe\ enhanced than early-type galaxies, only
because of their smaller masses. At a given velocity dispersion,
bulges and elliptical galaxies are indistinguishable as far as the
basic properties of their stellar populations are concerned.  No
significant correlations of the stellar population parameters with
Hubble Type as late as Sbc are found, instead. In other words, the
stars in bulges do not originate in the discs.  This result also
agrees with the finding that structural parameters like disc and bulge
scale lengths, as well as bulge-to-disc ratios, are correlated with
bulge luminosity rather than with Hubble type.

If central spheroids have the same properties in galaxies with and
without discs, this clearly favours inside-out galaxy formation
\citep{vdBosch98} according to which the disc forms after the
bulge. Models that aim to explain the formation of bulges through disc
fragmentation processes need to push the formation epoch to relatively
high redshifts assuming high dissipation efficiencies
\citep{Imeetal04}.  Only in spiral galaxies of Hubble types later than
Sbc discs can have a significant influence on the evolution of the
stellar populations in the bulge component. This fits with the fact
that Sersic index drops significantly in the transition between Hubble
types Sbc and Sd \citep{BGP05}. Secular evolution through the disc and
the phenomenon of pseudobulge formation is most likely restricted to
spirals of types Sc and later.

\section*{Acknowledgements}
Claudia Maraston, Robert Proctor, and Marc Sarzi are thanked for very
constructive discussions. The referee is acknowledged for the very
careful and constructive report. This project has been partly
supported by grant BMBF-LPD 9901/8-111 of the Deutsche Akademie der
Naturforscher Leopoldina.



\normalsize

\bsp
\label{lastpage}

\end{document}